\documentclass[10pt,letterpaper]{article}
\usepackage[top=0.85in,left=2.75in,footskip=0.75in]{geometry}

\usepackage{amsmath,amssymb}

\usepackage{changepage}

\usepackage[T1]{fontenc}
\usepackage[utf8x]{inputenc}

\usepackage{textcomp,marvosym}

\usepackage{cite}

\usepackage{nameref,hyperref}

\usepackage{microtype}
\DisableLigatures[f]{encoding = *, family = * }

\usepackage[table]{xcolor}

\usepackage{array}

\newcommand{\rev}[1]{{#1}}
\newcommand{\revb}[1]{{#1}}

\newcolumntype{+}{!{\vrule width 2pt}}

\newlength\savedwidth

\usepackage[aboveskip=1pt,labelfont=bf,labelsep=period,justification=raggedright,singlelinecheck=off]{caption}

\makeatletter
\renewcommand{\@biblabel}[1]{\quad#1.}
\makeatother

\usepackage{lastpage,fancyhdr,graphicx}
\usepackage{epstopdf}
\fancyheadoffset[L]{2.25in}
\fancyfootoffset[L]{2.25in}
\newcommand{\ie}{\textit{i.e.~}}

\newcommand{\etal}{\textit{et al.~}}

\graphicspath{{./images/}}

\begin{document}
	\vspace*{0.2in}
	\begin{flushleft}
		{\Large
			\textbf\newline{Large-scale simulations of biological cell sorting driven by differential adhesion follow diffusion-limited domain coalescence regime
		}}
		\newline
		\\
		Marc Durand\textsuperscript{1*}
		\\
		\bigskip
		\textbf{1} 
		Universit\'{e} de Paris, CNRS, UMR 7057, Mati\`{e}re et Syst\`{e}mes Complexes (MSC), Paris, France
		\\
		\bigskip

		*marc.durand@univ-paris-diderot.fr

\end{flushleft}
\section*{Abstract}
Cell sorting, whereby a heterogeneous cell mixture segregates and forms distinct homogeneous tissues, is one of the main collective cell behaviors at work during development. 
Although differences in interfacial energies are recognized to be a possible driving source for cell sorting, 
no clear consensus has emerged on the kinetic law of cell sorting driven by differential adhesion. 
 Using a modified Cellular Potts Model algorithm that 
allows for efficient simulations while preserving the connectivity of cells, we numerically explore cell-sorting dynamics over \rev{very} large scales in space and time. 
For a binary mixture of cells surrounded by a medium, increase of domain size follows a power-law with exponent $n=1/4$ independently of the mixture ratio, revealing that the kinetics is dominated by the diffusion and coalescence of rounded domains. 
\rev{We compare these results with recent numerical studies on cell sorting, and discuss the importance \revb{of algorithmic differences as well as boundary conditions} on the observed scaling.}

\section*{Author summary}
Cell sorting describes the spontaneous segregation of identical cells in biological tissues. This phenomenon is observed during development or organ regeneration in a variety of biological systems. Minimization of the total surface energy of a tissue, in which adhesion strengh between homotypic and heterotypic cells are different, is one of the mechanisms that explain cell sorting. This mechanism is then similar to the one that drives demixing of two immiscible fluids. Because of the high sensibility of this process to finite-size and finite-time effects, no clear consensus has emerged on the scaling law of cell sorting driven by differential adhesion. Using an efficient numerical code, we were able to investigate this scaling law on very large binary mixtures of cells. We show that on long times, cell sorting obeys a universal power law, which is independent of the mixture ratio.


\section*{Introduction}
Collective cell behaviors are involved in many morphogenetic events, such as embryo development, organ regeneration, wound healing, and the progression of metastatic cancer \cite{Alert_2020}.
One of the simplest and best studied examples of such behavior is the
spontaneous separation of two randomly mixed cell populations, in a process called cell sorting \cite{Foty_2013}.
In the 1960s, Steinberg
proposed the differential adhesion hypothesis (DAH) as a mechanism to explain cell sorting \cite{Steinberg_1962}. The DAH
postulates that the surface tension of tissue arises from cell adhesion, and that cell configurations
are determined by minimization of the surface energy of cell aggregates in a manner similar
to phase separation processes, although surface energy minimization is an active process for cell sorting: cellular activity is necessary to untrap from the many local minima of the rough energy landscape of a cellular system. Since Steinberg's hypothesis statement, several numerical studies have proven that the DAH could reproduce cell sorting phenomena similar to those observed experimentally \cite{graner_simulation_1992, glazier_simulation_1993, mombach_quantitative_1995, Mombach_rounding, Cochet-Escartin_2017}. 
However, no clear consensus has emerged from these studies on the kinetics law of \rev{cell sorting driven by differential adhesion}: in their seminal papers \cite{graner_simulation_1992, glazier_simulation_1993}, Graner and Glazier found -- on a somewhat smaller system -- that the \textit{boundary length}, defined as the number of mismatched cell contacts, decays logarithmically. More recently, Nakajima \etal \cite{nakajima_kinetics_2011} reported a power-law decay, with an exponent which depends on the proportion of the two cell types: from $n=1/3$ for even mixture to $n=1/4$ for uneven mixture. Cochet-Escartin \etal \cite{Cochet-Escartin_2017}  also reported power-law decays, but with significantly higher exponents: $n \in [0.49, 0.74]$ from experiments, while $n \in [0.55, 0.59]$ from simulations. 

\rev{Meanwhile, alternative mechanisms have been proposed in recent years to explain cell sorting phenomena -- such as differences in cell
motility that are either intrinsic \rev{\cite{Beatrici_2011}} or dependent on the
cells' local environment \cite{Kabla_2012, Strandkvist_2014}, or a combination of locally coherent motion with differential adhesion \cite{Belmonte_2008} -- leading to different scaling behaviours. Although differential adhesion is sufficient to explain cell segregation, several recent studies show that the tendency of cells to actively follow their neighbors reduces segregation
time scales \cite{Kabla_2012, Belmonte_2008, Brunnet_2017}.}

\rev{The kinetics of cell sorting is crucial for morphogenetic processes, which must proceed with well-defined order and
timing. Most importantly, characterizing the kinetics can help us to discriminate between the wide variety of proposed models, and thus would give us a deeper understanding of the
underlying mechanisms that induce cell sorting.
}
In this context, more extensive simulations of DAH-based cell sorting are needed to find out which type of kinetics law actually governs cell sorting. 
In this paper, we present extensive \rev{DAH-based} cell sorting simulations for two-dimensional binary mixtures of cells surrounded by medium, using a recently modified Cellular Potts Model (CPM) algorithm that allows for simulations on very large scales in space and time. On long times, reported kinetics obeys algebraic scaling law with exponent $n=1/4$ independently of the mixture ratio. We discuss the importance of \revb{algorithmic differences}, boundary conditions, initial cluster geometry, space dimension, and finite size effects on the observed scaling.

\section*{Methods}
\subsection*{Cellular Potts Model}

Cellular Potts model (CPM), also called the Glazier–Graner–Hogeweg (GGH) model, is one of the most accepted models of a multicellular system. It was initially created to simulate the behavior of single cells during cell sorting \rev{driven by differential adhesion}\cite{graner_simulation_1992, glazier_simulation_1993}. Since then, this model has been extended to reproduce many other collective behaviors of cells \cite{glazier_magnetization_2007, Vedula_2012, Boas_2018, Hirashima_2017}. 
Cellular activity is modeled with an effective temperature $T$, and the update algorithm is based on Metropolis-like dynamics \cite{mombach_quantitative_1995}.
The main justification for it is that average time evolution of a cell configuration then obeys the overdamped force-velocity relation $\mathbf F = \mu \mathbf{v}$, where $\mathbf v$ is the cell velocity, $\mu$ is an effective mobility, and $\mathbf F = - \nabla \mathcal H$ is the acting force  \cite{Glazier_2007}.
The CPM is a lattice based modeling technique: each cell is a subset of lattice sites sharing the same cell ID $\sigma$. A cell type $\tau(\sigma)$ is also defined for each cell. \rev{In our simulations cells are of two types, noted $\tau=B$ and $\tau=Y$, where $B$ and $Y$ stand for \textit{Blue} and \textit{Yellow}, the respective colors of the cells in our graphical outputs (see Fig \ref{fig:patterns}). To simulate a fluid medium that surrounds the cell aggregate, a third cell type $\tau=M$ is introduced (where $M$ stands for \textit{Medium}.)}
Following \cite{graner_simulation_1992, glazier_simulation_1993}, the CPM Hamiltonian $\mathcal{H}$ we use reads: 
\begin{equation}
	\mathcal{H}=\sum_{\substack{neighboring \\ sites  \langle k,l \rangle}}J_{\tau,\tau^\prime}\left(1-\delta_{\sigma_k,\sigma_l}\right)+\frac{B}{2A_0}\sum_{\substack{cells \\ i}}\left(A_i-A_0\right)^2.
	\label{Hamiltonian}
\end{equation}
The first sum in Eq. (\ref{Hamiltonian}) is carried over neighbouring sites $\langle k,l \rangle$ and represents the boundary energy: each pair of neighbours having unmatching indices determines a boundary and contributes to the boundary energy.
Here, $\sigma_k$ and $\sigma_l$ are the site values of site $k$ and $l$, respectively.
$\tau$ and $\tau^\prime$ are abbreviations for $\tau(\sigma_k)$ and $\tau(\sigma_l)$. 
 $J_{\tau,\tau^\prime}\left(=J_{\tau^\prime,\tau}\right)$ is the energy per unit contact length between cell types $\tau$ and $\tau^\prime$. 
 The second sum in Eq. (\ref{Hamiltonian}) represents the compressive energy of the cells. $B$ is the effective bulk modulus of a cell, which captures its out-of-plane elongation \cite{Hannezo_2014,Villemot_2020}, $A_i$ is the area of cell $i$, and $A_0$ the \textit{nominal area}.

Eq. \ref{Hamiltonian} is the minimal Hamiltonian whose minimization induces cell sorting phenomena. 
However, various additional terms can be added to account for real cell behaviors, or application of external fields \cite{ouchi_improving_2003,glazier_magnetization_2007,magno_biophysical_2015,Niculescu_2015}. \rev{Especially, it is common to add a perimeter penalty term to Eq. \ref{Hamiltonian} to include the effects of cortical tension \cite{nakajima_kinetics_2011, Glazier_2007, magno_biophysical_2015}. As this term does not depend on how the two cell types are disposed within the aggregate, it has no perceptible effect on the kinetics of cell sorting, which only requires that the adhesive energy of two heterotypic cells is less than that of two homotypic cells \cite{graner_simulation_1992, glazier_simulation_1993}.} 
\rev{Let us emphasize that, as we investigate kinetics of cell sorting driven by differential adhesion exclusively, we just use Eq. \ref{Hamiltonian} in our simulations; we do not add extra terms that would account for spatial or temporal correlations in cell motion \cite{Beatrici_2011,Kabla_2012,Strandkvist_2014,Belmonte_2008}.}

CPM presents a number of advantages over competing numerical models for multicellular systems, such as vertex- or voronoi-based models \cite{Osborne_2016}:  interfaces can have arbitrary shapes, T1 topological rearrangements or neighbor exchanges, which are crucial for dynamics in such systems, are naturally included,
 and free boundaries with the medium are handled as simply as boundaries between adjacent cells. However, a notorious weakness of CPM -- when used with the standard updating rule -- is that it does not guarantee connectivity of cells: as temperature increases, small fragments detach from cells, biasing the cell perimeters and cell centers of mass.
Ideally, CPM cell sorting simulations should be run in a temperature range such that interface fluctuations are high enough for generating T1 neighbors swapping, but low enough for avoiding cell fragmentation. However, such a temperature range is almost nonexistent: 
the thermal energy required to generate a T1 in a hexagonal pattern (honeycomb) is estimated to be a fraction of the typical energy of one side: $\Delta E_{T_1}\sim 0.1 \gamma L$ \cite{Durand_2019}, where $\gamma$ is the boundary energy per unit length. 
In comparison, the minimal energy to create a fragment with size $l$ is $\sim \gamma l$ \cite{durand_efficient_2016}. 
Then, fragments of size up to $l\simeq 0.1L$ show up before first T1 rearrangements occur.
To prevent the visual appearance of fragments, quenching process is usually performed before snapshots. However, the creation of fragments, inefficient for the cell sorting process, artificially slows down simulations \cite{durand_efficient_2016}.

\subsection*{\rev{Modified} algorithm that preserves cell connectivity}
In a recent work \cite{durand_efficient_2016}, we proposed a \rev{modified} CPM algorithm that forbids fragmentation and improves substantially computational efficiency for a same running temperature: the time wasted in checking cell connectivity is more than offset by the time spared in fragments moves, evidencing that fragmentation occurs even at low temperature. Moreover, efficiency can be further enhanced by working at temperature range that are unattainable with the standard algorithm, due to the proliferation of fragments. 
\rev{Like the standard CPM algorithm, this modified version follows a modified Metropolis updating rule: the value of a lattice site, chosen randomly (and called \emph{candidate site}), is changed to a value chosen randomly in its neighborhood (\emph{target value})) with probability $p=\min(1,e^{-\Delta E/T})$ (where $\Delta E$ is the change of energy associated with the modification), and under the extra-requirement that this change preserves the \emph{local connectivity} of the target and candidate cells. That is, all the sites with target (resp. candidate) value within the Moore neighborhood of the central site are adjacent to each other. Starting with an initial configuration where cells form simply connected domains, this test (which is performed at every copy attempt) ensures that connectivity of cells is preserved, \ie it prevents the formation of fragments or multiply connected cells.
Because our algorithm also belongs to the Metropolis-like class, we expect to observe the same kinetics than with the standard algorithm. Indeed, assuming that the frequency of fragmentation events that occur with the standard algorithm is homogeneous in time, the real time-correspondence of one MCS for both algorithms must be proportional to each other, for a given set of the different parameter values. 
}

\rev{In \cite{durand_efficient_2016}, we performed a few DAH-based cell sorting simulations on small systems to benchmark this modified algorithm and compare its efficiency with that of the standard CPM algorithm.}
Here we use this efficient algorithm to fully characterize \rev{DAH-based} cell sorting kinetics on unprecedentedly large systems. We performed simulations for binary mixtures of $N$ cells, with $N$ ranging from $200$ to $320,000$. All data were averaged over five independent runs. Moreover, in order to smooth out completely the anisotropy of the underlying square lattice, we used a 4th order neighborhood (composed of 20 lattice sites) in the evaluation of the Hamiltonian, whereas previous studies used a 2nd order neighborhood, composed of 8 lattice sites only \cite{graner_simulation_1992,glazier_simulation_1993,nakajima_kinetics_2011}. 
Consistently, we also used a larger cell size (the target area was set to $A_0=100~ \mathrm{pixels}$), to avoid interactions between opposite cell sides. Values of the other parameters are: $T=50$, $B=200$, $J_{BB}=J_{YY}=8$,  $J_{YB}=14$, $J_{YM}=10$, $J_{BM}=22$. \rev{The $J$'s thus satisfy the required inequalities for that energy minimization induces cell sorting \cite{graner_simulation_1992}}.

\section*{Results}
\begin{figure*}[htb]
	\centering
	\includegraphics[width=\linewidth]{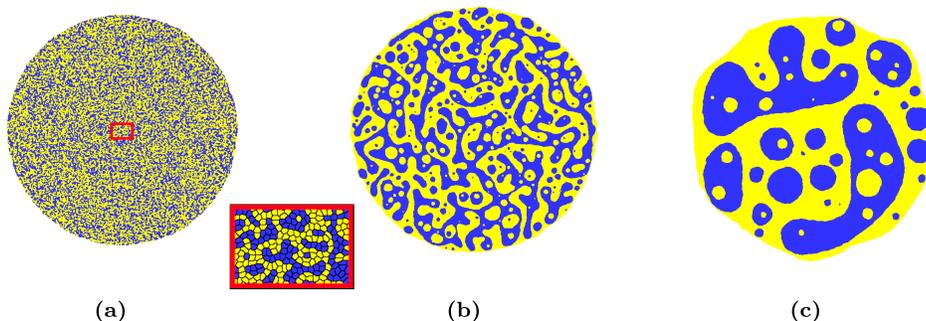}
	\caption{\label{fig:patterns} Time evolution of a cluster of $45\times 10^3$ cells with 50\%-50\% proportion of \rev{blue and yellow} cells: (a): t= $10^3$ MCS, (b): t= $40\times 10^3$ MCS, (c): t =$2\times10^6$ MCS. Close-up: zoom on the aggregate area indicated by a red rectangle. Nominal area of each cell is $100$ pixels$^2$.
}
\end{figure*}

Fig \ref{fig:patterns} shows different snapshots of the simulation for an even (50:50) mixture of cells. Initially, the two types of cells were distributed randomly in a rounded cluster (Fig  \ref{fig:patterns}A).
Subsequently, cells of the same type began to aggregate and form clusters (Fig \ref{fig:patterns}B), and the sizes of the clusters grew with time (Fig \ref{fig:patterns}C).

Choice of running temperature has a strong influence on the cell sorting timescale: fluctuations must be large enough to jump from a local minimum of energy to a lower one, but not too large to avoid jumping to a higher one, so there is likely an optimal temperature for which cell sorting timescale is minimal.
Most importantly, there is a full temperature range for which the \emph{kinetics} of cell sorting is preserved. Indeed, the net rate of transition from  configuration $i$ to configuration $j$ obtained with Metropolis-like algorithm is $r=1-e^{\Delta \mathcal H/T}\simeq  - \Delta \mathcal H/T$ for small relative change of energy $\Delta \mathcal H/T$. \rev{In this regime, a change in temperature uniformly scales the rate of transition between configurations, leaving the kinetics of cell sorting unchanged. Note that with the parameter values we used, $\Delta \mathcal H/T \simeq (J /T) \simeq 0.2$.}

\begin{figure}
	\centering
	\includegraphics[width=0.75\linewidth]{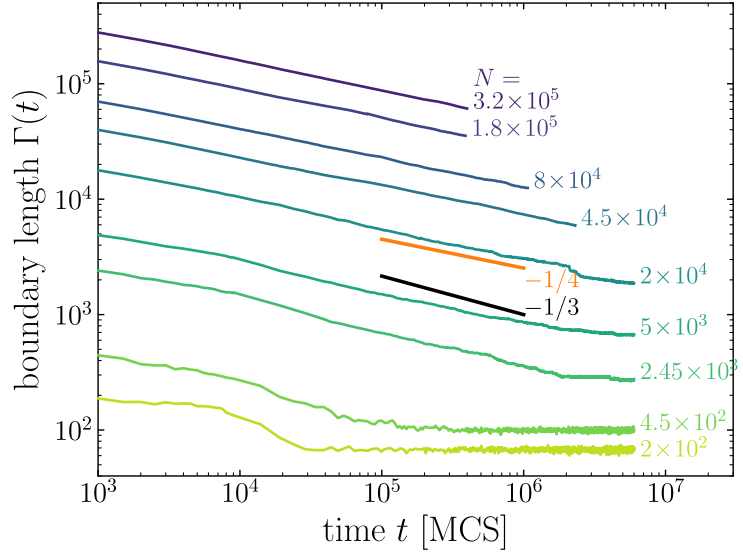}
	\caption{Time evolution of the boundary length $\Gamma(t)$ for \rev{even mixtures of $N$ cells}, with $N$ ranging from $200$ to $320,000$ (log-log scale). Power-law exponent is closer to $1/4$ than $1/3$. \label{fig:boundary-length}} 
\end{figure}

Following previous studies \cite{graner_simulation_1992, glazier_simulation_1993,Mombach_1995}, we quantify the kinetics of cell sorting by tracking the evolution of the \textit{total boundary length} $\Gamma(t)$, defined as the number of mismatched cell contacts. 
Fig \ref{fig:boundary-length} shows the evolution of $\Gamma(t)$ for even (50:50) mixtures of various sizes. We see that before complete cell sorting, \rev{curves for $N\geq 2\times 10^4$ cells follow the same algebraic variation: $\Gamma(t) \sim t^{-n}$  with $n \simeq 1/4$, attesting that this exponent is not affected by finite-size effects.} 
 Unexpectedly, our exponent value is different from the one reported by Nakajima and Ishihara  \cite{nakajima_kinetics_2011} \rev{for even mixture ratio} ($n=1/3$), but is identical to the value they obtain for uneven mixture ratios.

\begin{figure*}[htb]
	\centering
	\includegraphics[width=\linewidth]{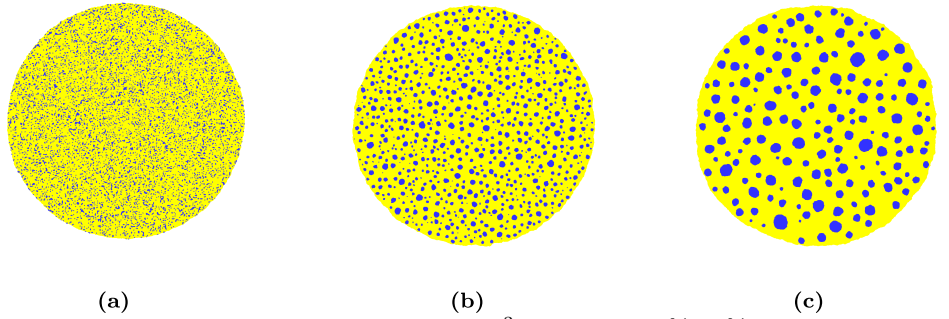}
	\caption{\label{fig:20-80-patterns} Time evolution of a cluster of $45\times 10^3$ cells with 20\%-80\% proportion of \rev{blue and yellow} cells: (a): t= $10^3$ MCS, (b): t=$40\times10^3$ MCS, (c): t =$2\times10^6$ MCS.
	}
\end{figure*}

\begin{figure}
	\centering
	\includegraphics[width=0.75\linewidth]{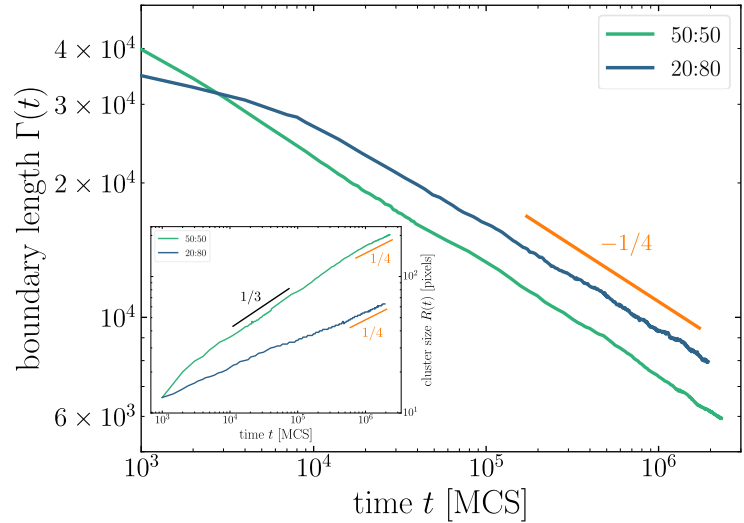}
	\caption{Time evolution of the boundary length $\Gamma(t)$ for even (50:50) and uneven (20:80) \rev{mixtures of $4.5\times 10^4$ cells (log-log scale).} Inset: time evolution of the cluster size $R(t)$ for the same systems.	\label{fig:boundary-length-20-80}}
\end{figure}

We then study how the mixture ratio affects the power-law exponent in our simulations. Fig \ref{fig:20-80-patterns} shows the typical time evolution of cell sorting for a 20:80 mixture. As for the 50:50 mixture, 
cells of the same type aggregate into clusters whose size grows with time. We note however that at same simulation time, clusters are much smaller and rounder than for the 50:50 mixture.
	Fig \ref{fig:boundary-length-20-80} compares the time evolution of boundary length for a 50:50 and a 20:80 mixture ratio. Both curves rapidly follow the same power law with exponent $n=1/4$. 


\begin{figure}[h]
	\centering
	\includegraphics[width=0.75\linewidth]{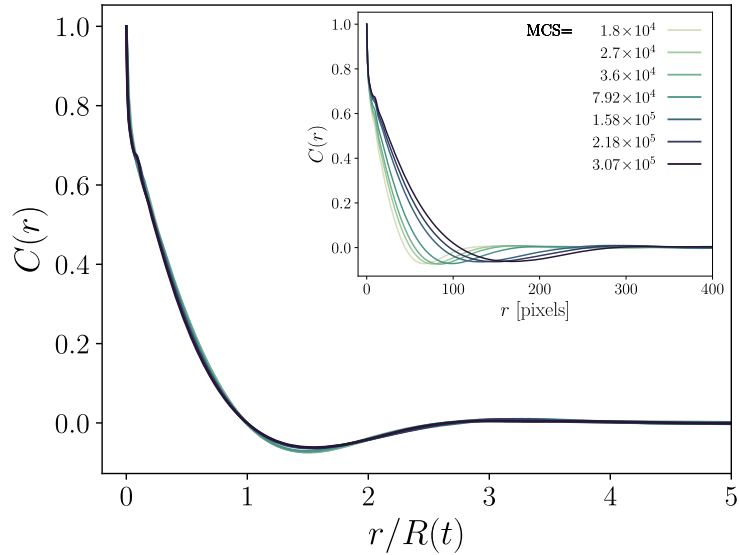}
	\caption{\rev{Autocorrelation} functions $C(r, t)$ as functions of the rescaled distance $r/R(t)$ for several time points, where $R(t)$ is determined as the position of the first zero of the raw \rev{autocorrelation} function at time $t$, shown in the inset.	\label{fig:correlation_function}}
\end{figure}

We also compare the evolution of the typical \textit{cluster size} $R(t)$ for the two mixture ratios. Following \cite{Huse_1986}, we define $R(t)$ as the position of the first zero of the \rev{autocorrelation} function
\begin{equation}
	C(\mathbf r,t)=\langle \sigma(\mathbf r_0,t)  \sigma(\mathbf r_0+\mathbf r,t) \rangle_{\mathbf r_0}-\langle \sigma(\mathbf r_0,t)   \rangle_{\mathbf r_0}^2,
\end{equation}
where $\sigma(\mathbf r_0,t)$ is the value at time $t$ of the lattice site located at $\mathbf r_0$, and $\langle \dots \rangle_{\mathbf r_0}$ denotes the average over all lattice site locations $\mathbf r_0$. 
Typical \rev{autocorrelation} functions are shown in the inset of Fig \ref{fig:correlation_function}. In agreement with \cite{nakajima_kinetics_2011}, when they are plotted with the rescaled distance $r/R(t)$, they all superimpose on a master curve, which shows that the system rescaled by $R(t)$ exhibits the same statistical property.
Note that boundary length and cluster size are related for circular clusters: $\Gamma(t) \sim N(t)R(t)$, where $N(t)$ is the number of clusters, whereas mass conservation implies $N\sim R^{-2}$, yielding $\Gamma \sim R^{-1}$. The evolution of $R(t)$ for the two mixture ratios is shown in the inset of Fig \ref{fig:boundary-length-20-80}. On long timescales, both curves follow the scaling regime $R(t)\sim t^m$ with $m=1/4$, indicating that clusters have circular shapes in this domain. Note however that for the uneven mixture this scaling regime is reached much more rapidly. For the even mixture, the curve follows a quite long transitory regime $R(t)\sim t^{1/3}$, indicating that clusters are not rounded yet.

Our results then suggest a universal scaling $n=m=1/4$ on long timescales, regardless of the mixture ratio, in contradiction with results of Nakajima and Ishihara \cite{nakajima_kinetics_2011}. 
Our exponent values also differ from those reported more recently by Cochet-Escartin \etal \cite{Cochet-Escartin_2017}. We discuss in the next section the origins of these discrepancies.

\section*{Discussion}
There are two major mechanisms 
for phase demixing and domain growth. 
The first one consists in the ``evaporation'' then ``condensation'' of the units (atoms, or here, cells) from the smaller clusters to the larger ones. In this \emph{evaporation-and-condensation} mechanism, centers of mass of clusters are fixed. 
The second mechanism is the diffusion then coalescence of clusters. In this \emph{diffusion-and-coalescence} mechanism, centers of mass of clusters are free to move. In real systems, combination of these two idealized mechanisms can occur.
For each of these two mechanisms, kinetics result from two competing processes: the growth rate of the clusters on one side, and the rounding of their interfaces on the other side.
The well known Ostwald ripening corresponds to evaporation-and-condensation mechanism when kinetics is limited by the transfer of material from smaller to larger clusters, which then have a rounded shape. Lifshitz, Sliozov and Wagner (LSW) theory then predicts that the domain growth follows a power law  $\sim t^{m}$ with exponent $m=1/3$ independently of space dimension \cite{Rogers_1989}. Although LSW theory assumes that the volume fraction of one of the components of the mixture is vanishingly small, it has been reported to hold for larger volume fraction as well \cite{Rogers_1989, Chakrabarti_1993}. 

As for the diffusion-and-coalescence mechanism, it is well-established \cite{Kolb_1984,Meakin_1990,Brunnet_2017} that in the diffusion-limited regime -- in which clusters have rounded shape -- domain growth also follows a power-law  with exponent $m=1/(2-d \alpha)$, where  $d$ is the space dimension and $\alpha$ the scaling factor between diffusing velocity and mass of aggregates. \rev{For cells with aligning interactions, the value of $\alpha$ is known to depend on the alignment strength \cite{Brunnet_2017}. For cells with no aligning interactions, as it is the case in our DAH simulations, $\alpha=-1$ \cite{nakajima_kinetics_2011, Brunnet_2017}}, yielding $m=1/4$ at two dimensions.

These results suggest that  in our simulations cell sorting is achieved through diffusion-and-coalescence mechanism and that on long times the kinetics is limited by the diffusion of (rounded) clusters, for both even and uneven mixture ratios. Actually, this should come as no surprise, since cluster diffusion has a slower kinetics than cluster rounding, which generally is characterized by an exponential rather than algebraic law \cite{Mombach_rounding, Cochet-Escartin_2017}.

At this point one question is left: how to explain the $n=1/3$ power-law reported by Nakajima and Ishihara for even mixture, while they did observe the $n=1/4$ regime for uneven mixture ? Although a $n=1/3$ power-law is characteristic of the LSW theory, this mechanism has been ruled out by Nakajima and Ishihara to explain the observed scaling, as the probability of detachment of two homotypic cells is vanishingly small with the parameter values they used. Note that with the parameter values we used, the probability of detachment is $p_\text{cell} = e^{−\Delta E_\text{cell}/T}$ with $\Delta E_\text{cell}  \sim (2J_{YB}-J_{YY}-J_{BB})z L \sim 7.5$, yielding $p_\text{cell}\sim5.5\times10^{-4}$. \revb{Hence, such events are unlikely in our simulations as well.} Besides, we also checked that detached cells are never observed in the simulation snapshots.

Rather than the detachment of homotypic cells, evaporation-and-condensation could proceed through the detachment of \emph{fragments} of cells. \revb{Obviously, such events do not occur with our non-fragmenting algorithm, but they are unavoidable with the standard CPM algorithm \cite{durand_efficient_2016}. Such a mechanism is particularly favored in even mixtures, in which the distance between homotypic clusters is minimized. That may explain why Nakajima and Ishihara observe the $n=1/3$ power-law for even mixtures, but not for uneven ones.}

\begin{figure}[h]
	\centering
	\includegraphics[width=0.7\linewidth]{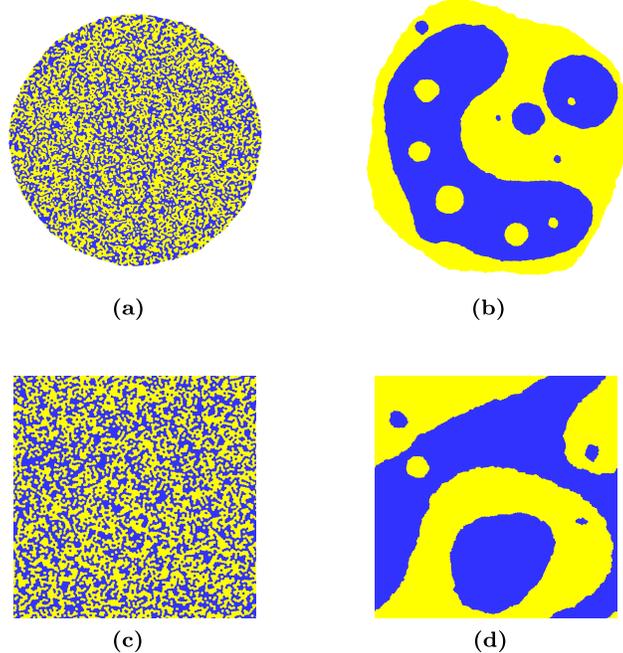}
	\caption{\label{fig:boundary-conditions}Time evolution of a cluster of $20\times 10^3$ cells with 50\%-50\% proportion of \rev{blue and yellow cells}: (a)-(b) cluster surrounded by medium (free boundary conditions) at $t=10^3$ MCS and $t=6\times 10^6$ MCS; (c)-(d) cluster with periodic boundary conditions at same time points.
	}
\end{figure}

\revb{
Apart from the use of a non-fragmenting algorithm, the other difference between both studies resides in the boundary conditions which are used: Nakajima and Ishihara used periodic boundary conditions, while in the present study, \emph{free} boundary conditions are used, meaning that cells are surrounded by the medium. 
This difference in boundary conditions, combined with finite size effects, could also explain why Nakajima and Ishihara do not observe the $n=1/4$ power-law for even mixtures. 
Indeed, as this is illustrated in Fig \ref{fig:patterns}B, there is a long transient regime in our simulations during which clusters form entangled structures in even mixtures (a point which is also discussed in \cite{Hutson_2008}): because of the high concentration of clusters, cluster growth process is faster than cluster rounding up process. The $m=1/3$ power-law variation of the cluster size $R(t)$ we observe (see inset of Fig \ref{fig:boundary-length-20-80}) corresponds to this transient regime where clusters grow in size but are not circular. As these clusters keep growing in size, their diffusion slows down. Eventually, on long times, rounding is faster and the kinetics follows the $n=1/4$ power-law, typical of the regime of diffusion and coalescence of circular clusters.
 When simulations are performed with periodic boundary conditions, these entangled structures rapidly span the simulation lattice and close on themselves via the periodic boundary conditions, as illustrated in Fig \ref{fig:boundary-conditions}. Such spanning clusters then stop diffusing immediately (as their displacements require a global move of the whole system), and slow down
 interface smoothing, because of a competition between spanning clusters that tend to form stripes with flat boundaries, and smaller clusters that tend to form circular boundaries. Nakajima and Ishihara attribute the observed $n=1/3$ power law to this interface smoothing process \cite{nakajima_kinetics_2011}. However, on long times, the $n=1/4$ diffusion process, which is slower than the smoothing process, should still prevail, unless there are no small diffusive clusters left because of the finite size of the simulated system. But then the mean cluster radius $R(t)$ should not evolve much as clusters stop growing, which is not what is reported in \cite{nakajima_kinetics_2011}.  
}

\begin{figure}[h]
	\centering
	\includegraphics[width=0.7\linewidth]{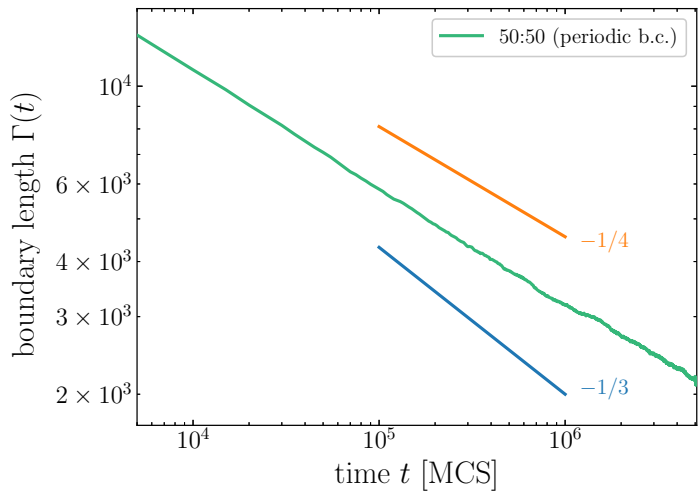}
	\caption{\label{fig:pbc}\revb{Time evolution of the boundary length $\Gamma(t)$ for even (50:50) mixture of $20\times 10^3$ cells with periodic boundary conditions.
	}
}
\end{figure}

\revb{
To get more insight on the exact cause of the discrepancy in power-law exponent, and discriminate between difference in algorithm and difference in boundary conditions, we performed cell sorting simulations of even mixtures with our modified CPM algorithm and using periodic boundary conditions. Contrary to Nakajima and Ishihara, we obtain a $n=1/4$ power-law for the evolution of boundary length (see Fig \ref{fig:pbc}), exactly as we obtained with the free boundary conditions. This result suggests that evaporation-and-condensation of cell fragments is at the origin of the $n=1/3$ power-law reported in \cite{nakajima_kinetics_2011}. Accordingly, the long-term scaling law is independent of boundary conditions and mixture ratio when using the non-fragmenting algorithm.
}

Initial geometry of the aggregate and dimension of space also affect the kinetics of cell sorting, as pointed out by Cochet-Escartin \textit{et al} \cite{Cochet-Escartin_2017}. In their study, cells were initially placed on a sheet in a 3D space. The exponents they observed in this configuration, either experimentally or numerically, were significantly higher than $1/4$. However, the initial flat configuration cannot affect the long time scaling of domain growth, and the evolution in a 3D space should rather decrease the exponent  (passing,  in the diffusion-limited regime, from $n=1/(2-d\alpha)=1/4$ in 2D to $n=1/5$ in 3D). Moreover, the exponents for $\Gamma(t)$ and $R(t)$ were quite different in their experimental data, suggesting that clusters did not have time to round up. Finite size effects, but also hydrodynamic couplings (for experiments) \cite{Farrell_1989, Hutson_2008}, could explain the larger exponent values they obtained.

The good agreement between the scaling law reported in the present study and in \cite{nakajima_kinetics_2011} for uneven binary mixtures confirms that the real time-correspondence of 1 Monte Carlo Step (MCS) for the standard CPM algorithm and ours, which prevents fragmentation, are proportional to each other $\tau'_{\text{1MCS}}=\alpha \tau_{\text{1MCS}}$, where the prefactor $\alpha$ depends only on the parameters that characterize the system (here $J_{ij}$, $B$ and $A_0$).
This linear relationship can be easily explained: as the frequency of fragmentation events that occur with the standard algorithm is homogeneous in time, the amount of time spent in the creation of fragments is homogeneous all along the sorting process.

\section*{Conclusion}
Using a modified CPM algorithm that forbids cell fragmentation and increases computational efficiency, we were able to assess the kinetics of cell sorting driven by differential adhesion on very large systems (up to $320,000$ cells). Our results show that on long times, kinetics is controlled by diffusion of rounded clusters binary mixture for cell aggregate surrounded by medium. The growth exponent is $n=1/4$ for two-dimensional systems, regardless of the mixture ratios of the two cell types.
\revb{Two explanations have been proposed to rationalize the apparent contradiction with exponent value reported previously: the use of a modified CPM algorithm which forbids cell fragmentation in one hand, and the use of different boundary conditions on the other hand. However, when using our non-fragmenting algorithm with periodic boundary conditions, we obtain the same $n=1/4$ power-law as with free boundary conditions, suggesting that the change in algorithm is at the origin of the difference in results. In order to confirm this result, further investigations must be conducted, with varying mixture ratio and simulation temperature. If this is verified, previous studies using the standard algorithm for simulating cell kinetics might have to be reconsidered, as they could be affected by the same nonrealistic cell fragmentation issue.} 

\rev{
As a concluding remark, we note that both $1/3$ and $1/4$ power laws are slower kinetics than those reported in recent experimental observations \cite{Cochet-Escartin_2017, Krieg_2008, Mehes_2012}. This, however, does not necessarily reject the differential adhesion hypothesis. Indeed, these exponents are obtained for ideal planar geometries, and in a steady regime, requiring systems with a large number of cells. These two assumptions are not easily fulfilled in experiments, and change of cluster geometry or finite size effects can alter the observed kinetics. These limitations will be taken into account in future research work. Other measures at the cell level, such as velocity distribution in the two cell populations or correlations in cell velocity can help to discriminate the different models. Using such complementary measures \cite{Cochet-Escartin_2017} to support the DAH over the differential motility model as the mechanism that drives regeneration of \emph{Hydra}. However, other experimental studies support collective cell migration as the main mechanism of cell segregation \cite{Mehes_2012}. It is therefore likely that cell sorting processes, which occur in a large variety of systems, cannot be explained by a single universal mechanism. Given that DAH-based cell sorting has slower kinetics, it must be involved in small multicellular systems primarily, while other mechanisms such as collective motility or chemotaxis might occur in larger systems.
}


\end{document}